\documentclass[twocolumn,showpacs,preprintnumbers,draftclsnofoot]{revtex4}
\usepackage{amsmath}
\usepackage{amssymb}
\usepackage{graphicx}
\usepackage{dcolumn}
\usepackage{bm}
\usepackage{amssymb}
\usepackage{graphicx}
\usepackage{dcolumn}
\usepackage{bm}

\begin{document}

\title{ Twisted bilayer graphene with Kekul$\acute{e}$ distortion: isolated flat band   }
\author{Ma Luo\footnote{Corresponding author:luom28@mail.sysu.edu.cn} }
\affiliation{The State Key Laboratory of Optoelectronic Materials and Technologies \\
School of Physics\\
Sun Yat-Sen University, Guangzhou, 510275, P.R. China}

\begin{abstract}

Twisted bilayer graphenes with magical angle exhibit strongly correlated electronic properties because of the isolated flat band at the Fermi level. We studied the twisted bilayer graphene with substrates on both layers. The substrate induce Kekul$\acute{e}$ distortion for each graphene layer. The systems are investigated by both continuous Dirac Fermion model and tight binding model. The two investigations give similar conclusion that isolated flat band with ultra-narrow bandwidth could appear.

\end{abstract}

\pacs{00.00.00, 00.00.00, 00.00.00, 00.00.00} \maketitle

\section{Introduction}

Twisted bilayer graphene (TBLG) has attracted a lot of attention because of the systems exhibit strongly correlated electronic physics, such as Mott insulator and superconductivity. The fact that superconductivity could be obtained in such simple system with only carbon element and tunable structure is extinguish. The realistic graphene could be described by tight binding model in honeycomb lattice with Hubbard interaction. The ratio between the Hubbard interaction strength $U$ and the nearest neighbor hopping strength $t$ is $U/t=1.6$ \cite{Schuler13}, so that the systems is not strongly correlated. For the bilayer graphene with a twist with magical angle \cite{LopesdosSantos07,EJMele10,Shallcross11,EJMele11,LopesdosSantos12,GuyTrambly16,Mingjian18,Ramires18,Adriana18,YoungWooChoi18,JianKang18,Yndurain19} , isolated flat bands  \cite{Morell10,Pilkyung13,Hirofumi17,Nguyen17,Angeli18,Naik18} near to the intrinsic Fermi level appear. The effective tight binding model for the flat band \cite{Venderbos18,Koshino18,Yuan18} have small nearest neighbor hopping term $t^{\prime}$, because the electrons are localized around the AA stacked region. Thus, large ratio of $U/t^{\prime}>5$ could be obtained, which drive the systems into strongly correlated electronic systems. The superconductivity in TBLG is confirmed in experiment \cite{YuanCao181,YuanCao182}. The topolotical nontrivial phase of TBLG is also found in experiment \cite{Spanton18,SSSunku18,JiaBinQiao18,Shengqiang18}. Some theoretical works have been devoted to explain this strongly correlated electronic systems \cite{ChengCheng18,Sherkunov18,Rademaker18,YuPingLin18,Peltonen18,HoiChun18,YingSu18,Kennes18,Fengcheng18,Cenke18,Hiroki18,Gonzalez19,Hejazi19,BitanRoy19}.

For a suspended TBLG, the atomic corrugation play important role in the band structure \cite{Lucignano19}. In order to enhance the mechanical stability of the TBLG and reduce the impact from atomic corrugation, we studied the bilayer graphene with substrates. The particular type of substrates with honeycomb lattice on the surface whose lattice constant is $\sqrt{3}a$ is considered, with $a$ being the lattice constant of graphene. An example of this type of substrate is $In_{2}Te_{2}$ \cite{Giovannetti15}. If the substrates are thick enough, the atomic corrugation of the TBLG layers could be neglected, because the strain is absorbed by the substrates. Kekul$\acute{e}$ distortion is induced in the graphene, whose primitive unit cell is $\sqrt{3}a\times\sqrt{3}a$ super-cell of the pristine graphene. In the presence of Kekul$\acute{e}$ distortion, the two Dirac cones are mixed together at the $\Gamma$ point of the Brillouin zone, and a gapped in open. The low energy states have quadratic band. The band gap could be tuned by changing the inter-layer distance between graphene and substrate, i.e. by pressure. The continuous Dirac Fermion model of the twist bilayer graphene with the Kekul$\acute{e}$ distortion is developed. The numerical results given by the continuous Dirac Fermion model and the tight binding model exhibit band structure with qualitatively similar feature. The flat bands near to the Fermi level could have band width smaller than 0.3 meV.

The article is organized as following: In section II, the lattice structure of the TBLG with Kekul$\acute{e}$ distortion is presence. In section III(A), the continuous Dirac Fermion model is described. In section III(B), the tight binding model is described. In section IV, the band structures given by the two models are investigated. In section V, the conclusion is given.

\section{The Lattice Structure}

\begin{figure}[tbp]
\scalebox{0.59}{\includegraphics{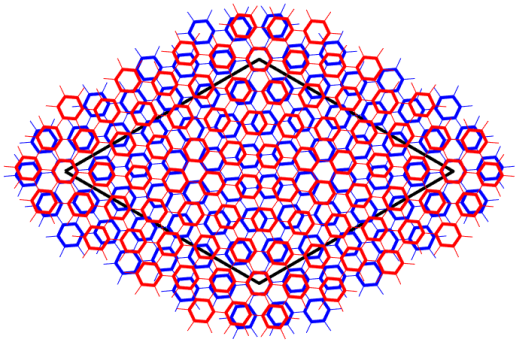}}
\caption{ The lattice structure of the TBLG with Kekul$\acute{e}$ distortion in each layer, and twisted angle being 9.43$^{o}$. The top and bottom layer are plotted as blue and red, respectively. The bond with Kekul$\acute{e}$ ratio is plotted as thick line. \label{fig1}}
\end{figure}

The scheme of constructing the superlattice of the twisted bilayer system is the same as that for the suspended twist bilayer graphene, except that the primitive unit cell is three time larger. For the single layer graphene with Kekul$\acute{e}$ distortion, the primitive unit cell include six carbon atoms, which are arranged in a hexagonal loop. The hexagonal loops are periodically arranged in a triangular lattice. The hopping between two carbon in the same hexagonal circle is changed by a Kekul$\acute{e}$ ratio designated as $\kappa$. The unit cell of the supperlattice includes three AA stacking regions. Thus, for the same twisted angle, the unit cell of the TBLG with Kekul$\acute{e}$ distortion is three times larger than that of the suspended bilayer graphene. The lattice structure of an example with twisted angle being 9.43$^{o}$ is plotted in Fig. \ref{fig1}.

\section{numerical model}

\subsection{Continuous Dirac Fermion Model}

\begin{figure}[tbp]
\scalebox{0.56}{\includegraphics{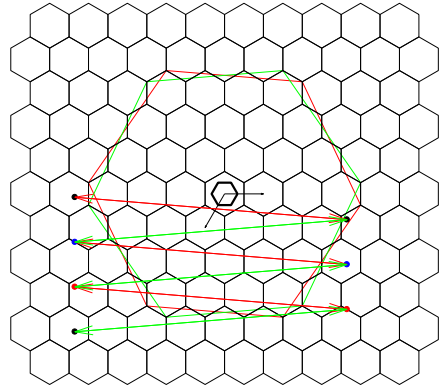}}
\caption{ The Brillouin zones of the TBLG. The first Brillouin zones of the upper and lower graphene layer are plotted as red and blue hexagons, respectively. The array of hexagons are the Brillouin zones of the suspended TBLG, with the black arrows being the two unit vectors $G_{M1}$ and $G_{M2}$. The hexagon with thick line in the middle of the figure is the first Brillouin zones of the TBLG with the Kekul$\acute{e}$ distortion. The black, blue and green dots stand for the three sets of nonequivalent wave vector in the Brillouin zones of the suspended twist bilayer graphene. The red and green arrows represent the inter-valley coupling due to the Kekul$\acute{e}$ distortion.   \label{fig2}}
\end{figure}

The continuous Dirac Fermion model of the TBLG with Kekul$\acute{e}$ distortion could be obtained by extending the continuous Dirac Fermion model for the suspended TBLG \cite{Koshino18}. The scheme in the reciprocal space are plotted in Fig. \ref{fig2}. The lattice vectors of the non-twisted AA stacking bilayer graphene are $\mathbf{a}_{1}=a\hat{x}$ and $\mathbf{a}_{2}=\frac{a}{2}\hat{x}+\frac{a\sqrt{3}}{2}\hat{y}$. The corresponding reciprocal lattice vectors are $\mathbf{a}_{1}^{*}$ and $\mathbf{a}_{2}^{*}$. The first Brillouin zones of the upper and lower graphene layer are twisted anticlockwise andclockwise for the angle $\theta/2$, respectively, by the rotation operator $R(\pm\theta/2)$. Thus, the lattice vectors of the $l$ layer are $\mathbf{a}_{1}^{(l)}=R(\pm\theta/2)\mathbf{a}_{1}$ and $\mathbf{a}_{2}^{(l)}=R(\pm\theta/2)\mathbf{a}_{2}$, with $\pm$ for $l=1,2$ standing for upper and lower layer, respectively. The corresponding reciprocal lattice vectors are then $\mathbf{a}_{1}^{(l)*}=R(\pm\theta/2)\mathbf{a}_{1}^{*}$ and $\mathbf{a}_{2}^{(l)*}=R(\pm\theta/2)\mathbf{a}_{2}^{*}$. The reciprocal lattice vectors for the suspended TBLG are $\mathbf{G}_{i}^{M}=\mathbf{a}_{i}^{(1)*}-\mathbf{a}_{i}^{(2)*}$ ($i=1,2$), which are plotted are black arrows in Fig. \ref{fig2}, and define the lattice of the Brillouin zones (hexagons in thin black line) of the suspended TBLG. In the presence of the Kekul$\acute{e}$ distortion, the unit cell is equal to the supercell of the $\sqrt{3}\times\sqrt{3}$ superlattice of the suspended TBLG, so that the first Brillouin zone is shrink in area for three times. The first Brillouin zone are plotted as hexagon in thick black line. For a wave vector inside the first Brillouin zone, the sampling wave vectors are obtained by adding integer numbers of $\mathbf{G}_{1}^{M}$ and $\mathbf{G}_{2}^{M}$, so that the wave vectors are in proximity to the $K$ and $K^{\prime}$ points of the non-twisted suspended bilayer graphene. The Kekul$\acute{e}$ distortion in each layer couples the wave vectors in $K$ and $K^{\prime}$ valleys with the same intra-valley wave vector, as shown by the red and green double arrows in Fig. \ref{fig2}. In details, for the upper layer, the wave vector $\mathbf{k}$ in the $K$ valley is coupled with the wave vector $\mathbf{k}+K^{\prime}_{(1)}-K_{(1)}$ in the $K^{\prime}$ valley, which is coupled to the wave vector $\mathbf{k}+K^{\prime}_{(1)}-K_{(1)}+K_{(2)}-K^{\prime}_{(2)}$ in the $K$ valley by the lower layer. Because $3(K^{\prime}_{(1)}-K_{(1)}+K_{(2)}-K^{\prime}_{(2)})=4\mathbf{G}_{1}^{M}+2\mathbf{G}_{2}^{M}$, after three rounds of inter-valley Kekul$\acute{e}$ coupling, the wave vector return to the original set of wave vectors for the suspended TBLG. As a result, the final set of wave vectors include the three inequivalent set of wave vector in both valleys, which form the low energy eigenstates. For each double arrows in Fig. \ref{fig2}, the coupling terms of $\Delta=(1-\kappa)t$ appear for the non-diagonal matrix element between different sublattices as well as different valleys \cite{ChangYuHou07}.

\subsection{Tight Binding Model}

The band structure of the magically strained bilayer graphene in the superlattice can be calculated by tight binding model. The Hamiltonian is given as
\begin{equation}
H=-\sum_{\langle i,j\rangle}{t_{i,j}c_{i}^{\dag}c_{j}}+c.c.
\end{equation}
, where $c_{i}^{(\dag)}$ is the annihilation (creation) operator of the electron at the $i$-th lattice cite, $t_{i,j}$ is the hopping parameter between the $i$-th and $j$-th lattice cites. The summation index cover the pairs of lattice cites $\langle i,j\rangle$, whose distance between each other is smaller than $5a_{c}$. The detail expression of $t_{i,j}$ could be found in multiple references \cite{Pilkyung13,Nguyen17,Adriana18,JianKang18}, which includes the intra-layer and inter-layer hopping parameter. For the hopping between two sites in the same hexagonal loop, the hopping strength is changed by the Kekul$\acute{e}$ ratio. Applying the Bloch boundary condition of the superlattice, the band structure of the supperlattice could be calculated. We define the figure of merit for the isolated narrow bands as
\begin{equation}
M=\frac{min(E_{gap}^{c},E_{gap}^{v})}{E_{w}}
\end{equation}
where $E_{gap}^{c(v)}$ is the gap from the conduction(valence) band, and $E_{w}$ is the band width of the narrow bands.

\section{The Band Structure}


\begin{figure}[tbp]
\scalebox{0.33}{\includegraphics{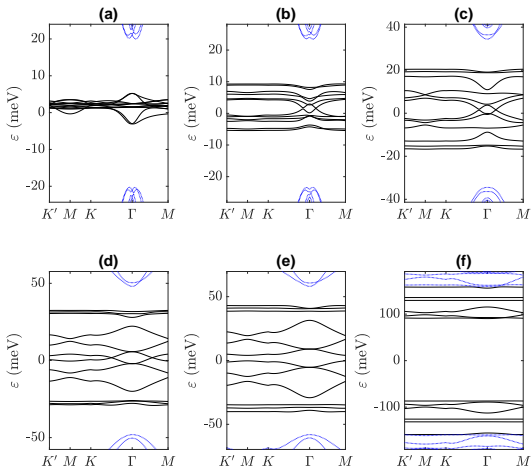}}
\caption{ The band structure of the TBLG given by the continuous Dirac Fermion model. The Kekul$\acute{e}$ parameter is (a) $\Delta=0$, (b) $\Delta=0.02$ eV, (c) $\Delta=0.05$ eV, (d) $\Delta=0.08$ eV, (e) $\Delta=0.1$ eV, (f) $\Delta=0.26$ eV. The twelve isolated bands around the Fermi level are plotted as black (solid) lines, and the bulk states are plotted as blue (dashed) lines.  \label{fig3}}
\end{figure}

\begin{figure}[tbp]
\scalebox{0.33}{\includegraphics{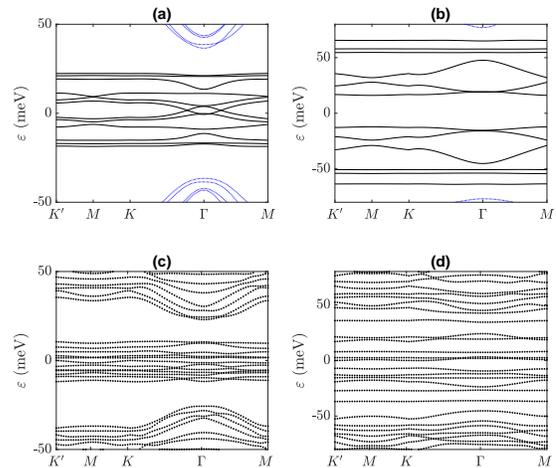}}
\caption{ The band structure of the TBLG given by the continuous Dirac Fermion model in (a-b), and by the tight binding model in (c-d). The Kekul$\acute{e}$ parameters in (a) and (c) are $\Delta=0.055$ eV, and in (b) and (d) are $\Delta=0.135$ eV.  \label{fig4}}
\end{figure}

The band structures of the TBLG with Kekul$\acute{e}$ distortion with magical angle 1.050$^{o}$ and varying Kekul$\acute{e}$ parameter are studied. By applying the continuous Dirac Fermion model, the band structures are plotted in Fig. \ref{fig3}. As $\Delta$ increase, the flat band around the Fermi level evolve. The band crossing points move and disappear. When $\Delta$ reach a large value, the bands are all very flat, and all band crossing or band degeneration point disappear. The band structures of the system with two Kekul$\acute{e}$ parameter are calculated by tight binding model, and the results are compared with the continuous Dirac Fermion model in Fig. \ref{fig4}. For the tight binding model, near nearest neighbor hopping are included, so that the particle-hole symmetric for each layer is broken, which in turn induces extra coupling effect. Therefore, the band crossing and band degeneration point become band avoid crossing. In addition, the band structures given by the tight binding model have flat bands closer to the Fermi level.

\begin{figure}[tbp]
\scalebox{0.33}{\includegraphics{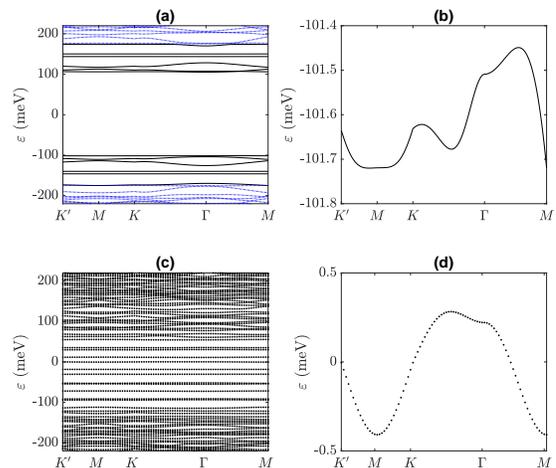}}
\caption{ The band structure of the TBLG given by the continuous Dirac Fermion model in (a-b), and by the tight binding model in (c-d). The figures (b) and (d) are zoom in of the figures (a) and (c) for the flat band near to the Fermi levels. The Kekul$\acute{e}$ parameter is $\Delta=0.28$ eV.    \label{fig5}}
\end{figure}

The band structures of a system with Kekul$\acute{e}$ parameter being $\Delta=0.28$ eV are plotted in Fig. \ref{fig5}. The flat band near to the Fermi level are zoomed in. As shown in Fig. \ref{fig5}(b) and (d), the detail structure of the flat band given by the two computational models have similar feature. the band width is smaller than 0.5 meV, so that the band is ultra flat. From the result of the tight binding model, the gap to the other flat band is about 10 meV, so that the figure of merit is larger than 20. Except for the flat band at the Fermi level, the other bands near to the Fermi level are also nearly flat band, so that the band structure near to the Fermi level is similar to the system with Landau levels. The phenomenon might be induced by the equivalent pseudomagnetic field in the twisted bilayer system \cite{JianpengLiu19}.

\section{conclusion}

In conclusion, TBLG with Kekul$\acute{e}$ distortion in each graphene layer host ultra-flat band with large gap from the higher and lower bands. A few higher and lower band near to the flat band at the Fermi level is also nearly flat. The figure of merit of the flat band is larger than 20. It would be interesting to study the model with interaction, which could exhibit the physics of strongly correlation and electron fractionalization \cite{ChangYuHou07,ClaudioChamon08,Bergman13,Bitan16}.

\begin{acknowledgments}
The project is supported by the National Natural Science Foundation of China (Grant:
11704419).
\end{acknowledgments}

\clearpage


\section*{References}

\begin{thebibliography}{99}


\bibitem{Schuler13} M. Schuler, M. Rosner, T. O. Wehling, A. I. Lichtenstein, and M. I. Katsnelson, Phys. Rev. Lett. 111, 036601(2013).






\bibitem{LopesdosSantos07} J. M. B. Lopes dos Santos, N. M. R. Peres, and A. H. Castro Neto, Phys. Rev. Lett. 99, 256802(2007).

\bibitem{EJMele10} E. J. Mele, Phys. Rev. B 81, 161405(R)(2010).
\bibitem{Shallcross11} S. Shallcross, S. Sharma, W. Landgraf, and O. Pankratov, Phys. Rev. B 83, 153402(2011).

\bibitem{EJMele11} E. J. Mele, Phys. Rev. B 84, 235439(2011).

\bibitem{LopesdosSantos12} J. M. B. Lopes dos Santos, N. M. R. Peres, and A. H. Castro Neto, Phys. Rev. B 86, 155449(2012).

\bibitem{GuyTrambly16} Guy Trambly de Laissardi$\grave{e}$re, Omid Faizy Namarvar, Didier Mayou, and Laurence Magaud, Phys. Rev. B 93, 235135(2016).

\bibitem{Mingjian18} Mingjian Wen, Stephen Carr, Shiang Fang, Efthimios Kaxiras, and Ellad B. Tadmor, Phys. Rev. B 98, 235404(2018).

\bibitem{Ramires18} Aline Ramires and Jose L. Lado, Phys. Rev. Lett. 121, 146801(2018).

\bibitem{Adriana18} Adriana Vela, M. V. O. Moutinho, F. J. Culchac, P. Venezuela, and Rodrigo B. Capaz, Phys. Rev. B 98, 155135(2018).

\bibitem{YoungWooChoi18} Young Woo Choi and Hyoung Joon Choi, Phys. Rev. B 98, 241412(R)(2018).

\bibitem{JianKang18} Jian Kang and Oskar Vafek, Phys. Rev. X 8, 031088(2018).

\bibitem{Yndurain19} Felix Yndurain, Phys. Rev. B 99, 045423(2019).




\bibitem{Morell10} E. Su$\acute{a}$rez Morell, J. D. Correa, P. Vargas, M. Pacheco, and Z. Barticevic, Phys. Rev. B 82, 121407(R)(2010).

\bibitem{Pilkyung13} Pilkyung Moon and Mikito Koshino, Phys. Rev. B 87, 205404(2013).

\bibitem{Hirofumi17} Hirofumi Nishi, Yu-ichiro Matsushita, and Atsushi Oshiyama, Phys. Rev. B 95, 085420(2017).

\bibitem{Nguyen17} Nguyen N. T. Nam and Mikito Koshino, Phys. Rev. B 96, 075311(2017).

\bibitem{Angeli18} M. Angeli, D. Mandelli, A. Valli, A. Amaricci, M. Capone, E. Tosatti, and M. Fabrizio, Phys. Rev. B 98, 235137(2018).

\bibitem{Naik18} Mit H. Naik and Manish Jain, Phys. Rev. Lett. 121, 266401(2018).






\bibitem{Venderbos18} J$\ddot{o}$rn W. F. Venderbos and Rafael M. Fernandes, Phys. Rev. B 98, 245103(2018).

\bibitem{Koshino18} Mikito Koshino, Noah F. Q. Yuan, Takashi Koretsune, Masayuki Ochi, Kazuhiko Kuroki, and Liang Fu, Phys. Rev. X 8, 031087(2018).

\bibitem{Yuan18} Noah F. Q. Yuan and Liang Fu, Phys. Rev. B 98, 045103(2018)






\bibitem{YuanCao181} Yuan Cao, Valla Fatemi, Shiang Fang, Kenji Watanabe, Takashi Taniguchi, Efthimios Kaxiras and Pablo Jarillo-Herrero, Nature, 556, 43每50(2018).

\bibitem{YuanCao182} Yuan Cao, Valla Fatemi, Ahmet Demir, Shiang Fang, Spencer L. Tomarken, Jason Y. Luo, Javier D. Sanchez-Yamagishi, Kenji Watanabe, Takashi Taniguchi, Efthimios Kaxiras, Ray C. Ashoori and Pablo Jarillo-Herrero, Nature, 556, 80每84(2018).

\bibitem{Spanton18} Eric M. Spanton, Alexander A. Zibrov, Haoxin Zhou, Takashi Taniguchi, Kenji Watanabe, Michael P. Zaletel, Andrea F. Young, Science, 360, 62每66(2018)

\bibitem{SSSunku18} S. S. Sunku, G. X. Ni, B. Y. Jiang, H. Yoo, A. Sternbach, A. S. McLeod, T. Stauber, L. Xiong, T. Taniguchi, K. Watanabe, P. Kim, M. M. Fogler, D. N. Basov, Science, 362, 1153每1156(2018).

\bibitem{JiaBinQiao18} Jia-Bin Qiao, Long-Jing Yin, and Lin He, Phys. Rev. B 98, 235402(2018).

\bibitem{Shengqiang18} Shengqiang Huang, Kyounghwan Kim, Dmitry K. Efimkin, Timothy Lovorn, Takashi Taniguchi, Kenji Watanabe, Allan H. MacDonald, Emanuel Tutuc, and Brian J. LeRoy, Phys. Rev. Lett. 121, 037702(2018).






\bibitem{ChengCheng18} Cheng-Cheng Liu, Li-Da Zhang, Wei-Qiang Chen, and Fan Yang, Phys. Rev. Lett. 121, 217001(2018).

\bibitem{Sherkunov18} Yury Sherkunov and Joseph J. Betouras, Phys. Rev. B 98, 205151(2018).

\bibitem{Rademaker18} Louk Rademaker and Paula Mellado, Phys. Rev. B 98, 235158(2018).

\bibitem{YuPingLin18} Yu-Ping Lin and Rahul M. Nandkishore,
Phys. Rev. B 98, 214521(2018).
%

\bibitem{Peltonen18} Teemu J. Peltonen, Risto Ojaj$\ddot{a}$rvi, and Tero T. Heikkil$\ddot{a}$, Phys. Rev. B 98, 220504(R)(2018).

\bibitem{HoiChun18} Hoi Chun Po, Liujun Zou, Ashvin Vishwanath, and T. Senthil, Phys. Rev. X 8, 031089(2018).

\bibitem{YingSu18} Ying Su and Shi-Zeng Lin, Phys. Rev. B 98, 195101(2018).

\bibitem{Kennes18} Dante M. Kennes, Johannes Lischner, and Christoph Karrasch, Phys. Rev. B 98, 241407(R)(2018).

\bibitem{Fengcheng18} Fengcheng Wu, A. H. MacDonald, and Ivar Martin, Phys. Rev. Lett. 121, 257001(2018).

\bibitem{Cenke18} Cenke Xu and Leon Balents,
Phys. Rev. Lett. 121, 087001(2018).

\bibitem{Hiroki18} Hiroki Isobe, Noah F.Q. Yuan, and Liang Fu, Phys. Rev. X 8, 041041(2018).

\bibitem{Gonzalez19} J. Gonz$\acute{a}$lez and T. Stauber, Phys. Rev. Lett. 122, 026801(2019).
%

\bibitem{Hejazi19} Kasra Hejazi, Chunxiao Liu, Hassan Shapourian, Xiao Chen, and Leon Balents, Phys. Rev. B 99, 035111(2019).

\bibitem{BitanRoy19} Bitan Roy and Vladimir Juri$\check{c}$i$\acute{c}$, Phys. Rev. B 99, 121407(R)(2019).





\bibitem{Lucignano19} Procolo Lucignano, Dario Alf$\grave{e}$, Vittorio Cataudella, Domenico Ninno, and Giovanni Cantele, Phys. Rev. B 99, 195419(2019).

\bibitem{Giovannetti15} Gianluca Giovannetti, Massimo Capone, Jeroen van den Brink, and Carmine Ortix, Phys. Rev. B 91, 121417(R)(2015).

\bibitem{JianpengLiu19} Jianpeng Liu, Junwei Liu, and Xi Dai, Phys. Rev. B 99, 155415(2019).



\bibitem{ChangYuHou07} Chang-Yu Hou, Claudio Chamon, and Christopher Mudry, Phys. Rev. Lett. 98, 186809(2007).

\bibitem{ClaudioChamon08} Claudio Chamon, Chang-Yu Hou, Roman Jackiw, Christopher Mudry, So-Young Pi, and Gordon Semenoff, Phys. Rev. B 77, 235431(2008).

\bibitem{Bergman13} Doron L. Bergman, Phys. Rev. B 87, 035422(2013).

\bibitem{Bitan16} Bitan Roy and Igor F. Herbut, Phys. Rev. B 93, 155415(2016).





\end{thebibliography}
\end{document}